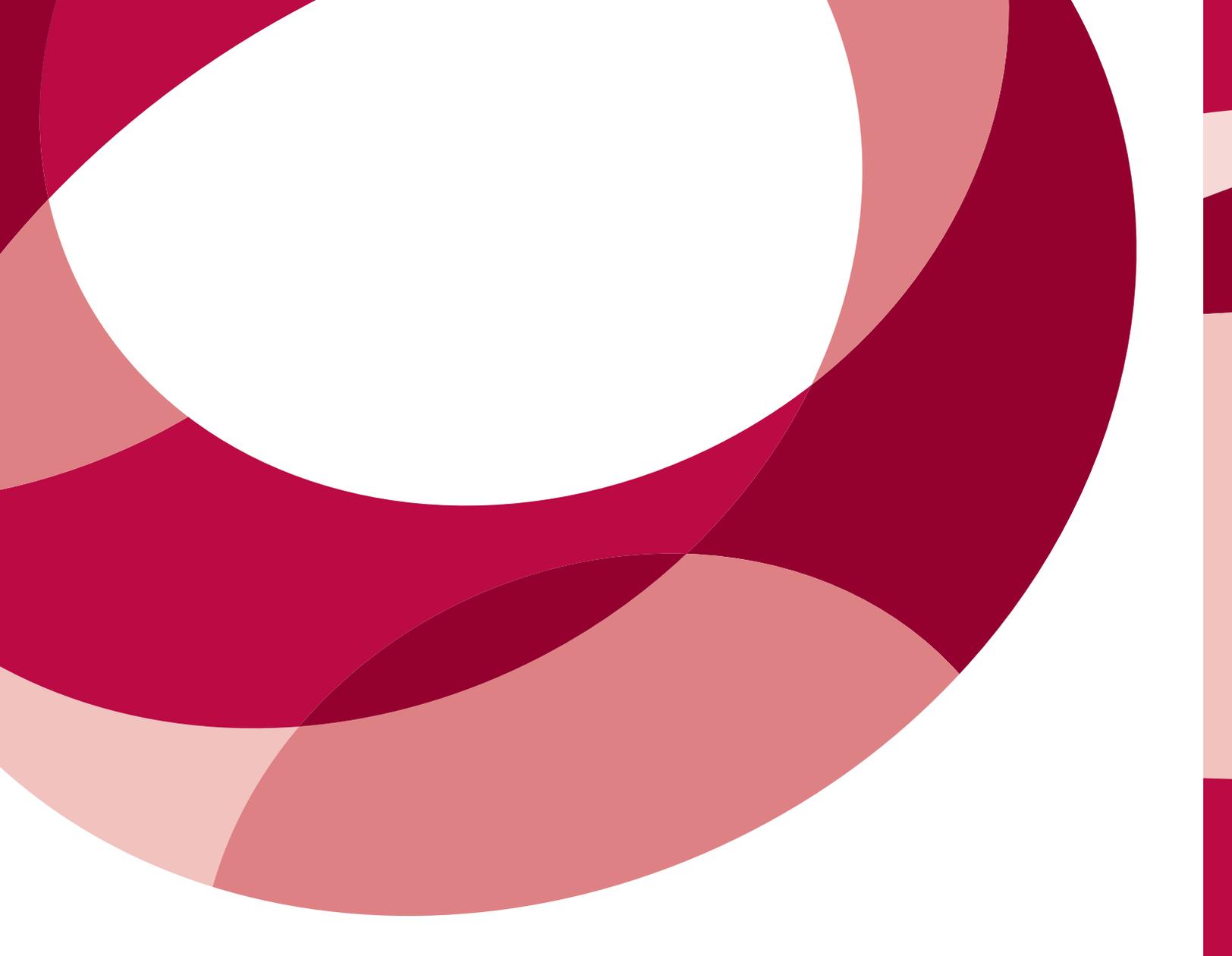

# Computer-Aided Personalized Education

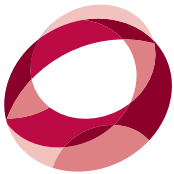

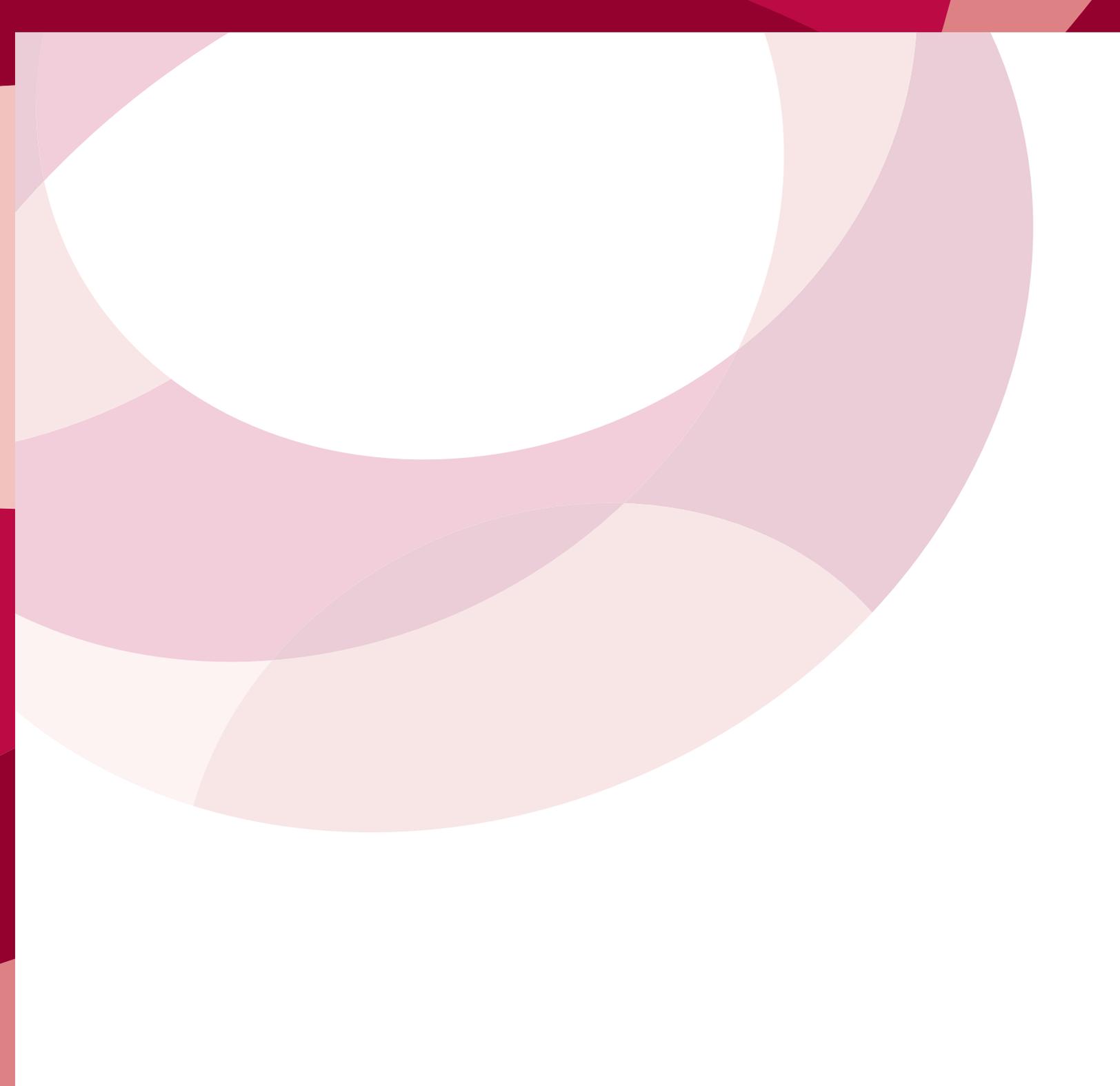

This material is based upon work supported by the National Science Foundation under Grant No. 1136993. Any opinions, findings, and conclusions or recommendations expressed in this material are those of the author(s) and do not necessarily reflect the views of the National Science Foundation.

# Computer-Aided Personalized Education

Rajeev Alur, Richard Baraniuk, Rastislav Bodik, Ann Drobnis, Sumit Gulwani,
Bjoern Hartmann, Yasmin Kafai, Jeff Karpicke, Ran Libeskind-Hadas, Debra Richardson,
Armando Solar-Lezama, Candace Thille, Moshe Vardi

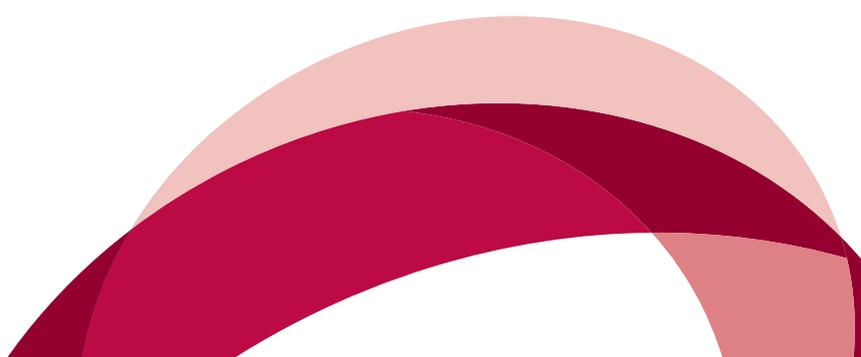

# 1. Introduction

The shortage of people trained in STEM fields is becoming acute. According to a recent study, there are 2.5 entry-level job postings for each new four-year graduate in STEM (see www.burning-glass.com/research/stem). Universities and colleges are straining to satisfy this demand. In the case of computer science, for instance, the number of US students taking introductory courses has grown three-fold in the past decade. Recently, massive open online courses (MOOCs) have been promoted as a way to ease this strain. This at best provides access to education. The bigger challenge though is coping with heterogeneous backgrounds of different students, retention, providing feedback, and assessment. Personalized education relying on computational tools can address this challenge.

While automated tutoring has been studied at different times in different communities, recent advances in computing and education technology offer exciting opportunities to transform the manner in which students learn. In particular, at least three trends are significant. First, progress in logical reasoning, data analytics, and natural language processing has led to tutoring tools for automatic assessment, personalized instruction including targeted feedback, and adaptive content generation for a variety of subjects. Second, research in the science of learning and human-computer interaction is leading to a better understanding of how different students learn, when and what types of interventions are effective for different instructional goals, and how to measure the success of educational tools. Finally, the recent emergence of online education platforms, both in academia and industry, is leading to new opportunities for the development of a shared infrastructure to facilitate large-scale deployment of educational tools for data sharing and experimentation. To articulate a long-term research agenda for transforming the technology for personalized education building on these trends, this CCC workshop brought together researchers developing educational tools based on technologies such as logical reasoning and machine learning with researchers in education, human-computer interaction, and cognitive psychology.

The scope of this report is focused primarily at college-level STEM subjects, including computer science, but with the understanding that training of high school students in these topics is essential to success. We begin with a survey of the emerging trends in personalized education tools and science of learning in section two. In section three, we outline a collective vision of how technology can transform learning, and conclude with research challenges to achieve this vision in section four.

# 2. Emerging Trends

In section two, we focus on problems central to computer-aided personalized education: formalization of tasks such as assessment, feedback, and content generation as computational problems, algorithmic tools to solve the resulting problems at scale, and effective integration of these tools in learning environments. Below we summarize recent trends in different disciplines aimed at solving these problems.

**Logical Reasoning**

In the last two decades, advances in automated reasoning tools such as model checkers and constraint solvers have led to successful applications to industrial scale software systems. A more recent application of logical reasoning is program synthesis – automatic derivation of a program from its high-level specification. Emerging research has shown that reasoning tools developed for verification and synthesis can be effectively used to solve computational problems in personalized education.

To understand the role of logical reasoning in personalized education, consider the task of automatically evaluating a student's submission to a programming problem in an introductory programming course. A commonly used assessment technique is to execute the student's program on a suitably chosen set of test inputs and check whether the resulting outputs match the expected ones. If this is not the case, instead of simply showing an input on which the program did not work correctly, a reasoning tool can try to synthesize a variant of the student's program that works correctly. The edits that are needed to obtain such a correction are then used to highlight lines of code that need to be changed or to provide hints. The tool AutoProf [SGS13] implements this strategy by relying on state-of-the-art tools for verification and synthesis, and its effectiveness has been demonstrated in evaluating students' submissions in introductory programming course at MIT.

Tools rooted in logical reasoning for tasks such as automatic generation of problems of a certain difficulty level, automatic grading, and automatic generation of hints have been developed for problems arising in a diverse set of computer science courses such as Algorithms, Automata Theory, Compilers, Databases, Programming, and Embedded Systems [Gul14,JDJS14,DK+15].

**Machine Learning**

Algorithms for machine learning are also beginning to move from industry into education. Current applications range from learning analytics tools that help students and instructors keep track of learning progress to personalized feedback tools that recommend the next best learning activity for a student based on their activities and progress to date. An example of such a personalized feedback tool can be found in [MXAS16], where the system can provide personalized predictions of a student's comprehension and predict his/her grade in the class. If the student performance in a class is low, the student is referred to an artificial intelligence system, called e-Tutor, which provides automatic remedial help that is personalized to the student, and has been shown to be helpful in large undergraduate classes [TBS15].

In contrast to logical reasoning approaches, machine-learning analytics typically eschew domain-specific models in favor statistical models trained from large amounts of student data.

As an illustrative example of this approach, consider the Sparse Factor Analysis (SPARFA) framework [LWSB14] which mines student grade book data to learn the latent concepts that underlie a subject. Once these concepts have been identified, SPARFA can assess a student's mastery of the concepts and track it over time to provide useful feedback to both the student and instructor. SPARFA can also autonomously organize the subject's course content (lecture notes, homework problems, feedback hints) by building a graph connecting those items to latent concepts. This toolset is integrated into the free, open source Openstax College textbooks (see www.openstaxcollege.org).

Another application of machine learning, clustering based on syntactic features of a student's solution has been used to identify the higher level strategy used by the student and match it with the feedback provided by a teacher for that strategy. Clustering techniques have been used for power-grading of short answer questions and mathematical calculations by grouping responses into different buckets [BBJV14, NPHG14, LVWB15].

**Student-computer Interaction**

A key challenge in the design of an effective personalized education environment is to allow the student to interact in a natural and intuitive manner. Researchers in natural language processing and human-computer interaction are increasingly developing tools and techniques to address this challenge.

Examples of applications of natural language processing (NLP) technology to personalized education include automatic generation of questions related to factual content in new subject matter, support for group processes in scientific reasoning tasks, and automatic grading of essays [RV05, BBV12].

Recent advances in computer graphics and virtual reality technology offers rich possibilities for gamification of education that can motivate students to learn new concepts via games. As a concrete example, consider Crystallize, an immersive collaborative game for second language learning [CA+16]. Since humans comprehend linguistic meaning through concrete experiences situated in the real world, becoming fluent in a new language in a classroom setting is difficult. In this 3D game, players navigate a virtual environment that simulates being immersed in a real target language environment. Players collaborate through language quests that require them to find words in the environment needed to accomplish objectives. Both contextual information on use of language and collaboration have been shown to dramatically improve learning outcomes.

The role visualization can play in learning is evident from the success of Python Tutor. It uses visual interactions to help people overcome a fundamental barrier in learning programming, namely, understanding what happens as the program executes different lines of code, and is being used by millions of people [Guo13] (see also www.pythontutor.com).

**Learning Science**

Cognitive science aims to understand some of the principles and processes involved in learning. The natural question then is, how can we use computational tools to support these processes? Recent years have witnessed increasing research in constructing mental models of students based on their interaction with educational tools, data mining past history of interactions to suggest next steps, experimental analysis of how learning outcomes are impacted by interventions, and understanding of the role of social factors in learning.

As an example of how technology can help learning outcomes, consider the problem of detecting whether or not a student is attentive while either sitting in a lecture, reading a book, or interacting with an online tool. Sensor technology and smart cameras can now detect wandering minds with high fidelity [KDM15]. Such technologies are leading to interactive books with a huge potential of impacting education.

Learning science tells us that students learn best when they have an opportunity to collaborate, discuss, and form communities. This has been already put in practice in supportive collaborative learning environments such chat rooms in MOOCs and similar platforms [AD+14,CL+15].

## 3. Vision

Researchers from different disciplines have demonstrated the benefits of personalized education tools in specific courses. We can build on this momentum and bring together researchers with different expertise in large-scale projects aimed at transformative changes in education technology. We envision that progress in personalized education technology can benefit the society in following ways:

◗ Our goal is to train students in advanced topics without having them sacrifice quality of life. This can be achieved by improving effectiveness of education through technology by maximizing learning at realistic time investments by teachers as well as students. One concrete measure of success could be that a future sophomore computer science student will know what a current senior student knows.

◗ Current techniques for assessment are focused on short-term learning. We envision a future of personalized learning apps that stimulate and incentivize people to be lifelong learners with a focus on long-term learning and knowledge retrieval on demand.

◗ A key challenge to personalized education is to foster a robust pipeline of a diverse group of students to STEM and related disciplines. Personalization can meet the demands of heterogeneous backgrounds and different learning styles, and ensure engagement and retention.

## 4. Research Roadmap

To realize our vision of how computer-aided personalized education technology can impact society, we need to make progress on the following research goals. We first list some long-term projects that will require sustained collaboration among computational and learning scientists.

◗ Current personal tutors are invariably focused on specific concepts in individual courses.

A ten-year goal is to develop an expert teacher per computer science student. Such a personalized assistant can track an individual student's progress throughout the curriculum by actively providing feedback and help.

◗ A comprehensive theory of learning is an achievable ten-year target. Such a theory can in turn impact personalized teachers by constructing a mental model of the student, adapting to how a student is responding to interventions, and accounting for social factors in learning such as collaboration.

◗ A key to progress will be the availability of shared large-scale data repositories and experimental testbeds to evaluate research ideas. Building such open-source and shared infrastructure is itself a challenging, long-term, and worthy research goal.

To conclude this report, we list promising topics that can be explored by research teams. Progress on these topics in the next few years can provide the building blocks necessary to achieve the long-term potential of personalized education.

◗ Scalability: Many educational tasks such as feedback generation can be cast as search problems in a large space of candidate artifacts. On one hand, there has been significant research and engineering investments in generic search technologies such as SAT and SMT solvers. While these techniques work fairly well for certain domains and small problem instances, they do not constitute a universal scalable solution. On the other hand, domain-specific search techniques that leverage knowledge of the underlying domain scale well in the target domain, but require significant time, research expertise, and engineering effort. An important future research direction is to enable easy construction of search techniques that scale well to various domains by integrating generic with domain-specific components.

◗ Mental models for feedback: Tools today are designed to give feedback based only on the student submission, but not so much based on the mental model of a student. Various forms of feedback are possible, and the one that should be presented to a student should be based on some modeling of the state of the student, such as learning style, past knowledge, and understanding of certain concepts. While such models have been studied in cognitive science, their incorporation in computational feedback tools is a pressing and challenging problem.

◗ Beyond STEM courses: Current tools for problem generation, automatic grading, and feedback generation focus on mathematical problems in STEM subjects. Such problems are amenable to computational formalization (a notable exception is grading of essays for grammar and style). Developing techniques that are more broadly applicable will require novel integration of many approaches, and offers a promising research opportunity.

◗ Multi-modal interfaces: Rapid advances in sensor technology are leading to new ways in which a human can interact with a computer, such as by text, by speech, and by touch. Such natural modes of interaction are particularly relevant for student engagement. At the same time, the specific goals of such tools can help alleviate challenges in computationally difficult tasks like translating natural language to a formal language. Thus, developing effective multi-modal interfaces for personalized education tools is an opportunity for creative research at the intersection of many disciplines.

◗ Collaborative learning environments: There is plenty of empirical evidence that students learn by collaborating with one another. Tools such as chat rooms and peer grading have been incorporated in current online learning environments. However, to fully achieve the promise of collaboration, we need research for better understanding of principles of both the role of collaboration in learning and how to add collaboration to learning.

◗ Predictive models: One important goal for a personalized education tool is to help struggling students meet their educational goals. Predictive models based on modern data analytics can detect potential problems in advance. For example, grades in quizzes early in a course can reliably predict the final course grade. An interesting research question is, how to design intervention strategies based on such predications, integrate them in learning environments, and ensure tangible improvements in learning outcomes?

◗ Adaptive syllabi and curricula: In a typical course, whether in a classroom or online, the content of the course is fixed in advance. Adaptive learning technology offers exciting opportunities to make the content dynamic. At the micro level, there is already some success in using computational tools for problem generation for specific concepts to suggest the next problem based on the student's past interactions. New research though is needed for adaptive content generation to dynamically develop the sequence of concepts resulting in a course that meets the desired learning outcomes and the sequence of courses leading to a curriculum that meets the desired breadth and depth requirements.

◗ Virtual Labs: A central component of engineering education is learning by building artifacts in a lab. This raises the question: can we create online labs with a learning experience close to the physical lab. Virtual simulation environments integrated with learning technology can offer a solution, and this leads to a number of research questions.

◗ Long-term learning outcomes: Traditionally testing is used to assess how much the student has learned during a course. A more meaningful assessment would be to measure how much knowledge a student retains over a long period and whether this knowledge can be retrieved as needed to solve problems. Cognitive science helps us understand how humans store and retrieve knowledge. A fruitful research direction is to integrate this understanding in personalized education tools to improve long-term learning outcomes.

◗ Privacy: Tools for personalized education base their decisions on mining data from students' solutions and students' history of interactions. These decisions cannot be made without access to sensitive information, but naturally lead to concerns about preserving privacy of individual students. Since personalized education is a nascent technology, it would be prudent to bake privacy concerns into tools right from the beginning. Finding the balance between information access and privacy and enforcement mechanisms are challenging technical problems, and research is needed to find solutions appropriate for the domain of education.

If we follow these suggestions as a community, we will make significant progress towards not only better educating STEM students with diverse backgrounds but also great strides in creating educational tools that will impact all students as we realize the benefits of personalized education.

## Participants

- Rajeev Alur, University of Pennsylvania
- Nina Amla, National Science Foundation
- Erik Andersen, Cornell University
- Anindya Banerjee, National Science Foundation
- Richard Baraniuk, Rice University
- Lida Beninson, National Science Foundation
- Gautam Biswas, Vanderbilt University
- Rastislav Bodik, University of Washington
- Emma Brunskill, Carnegie Mellon University
- Andy Butler, University of Texas, Austin
- Isaac Chuang, Massachussets Institute of Technology
- Sandra Corbett, Computing Research Association
- Loris D'Antoni, University of Wisconsin, Madison
- Lucas de Alfaro, University of California, Santa Cruz
- Sidney D'Mello, Notre Dame University
- Khari Douglas, Computing Community Consortium
- Ann Drobnis, Computing Community Consortium
- Barbara Ericson, Georgia Institute of Technology
- Kathi Fisler, Worcester Polytechnic Institute
- Michael Gleicher, University of Wisconsin, Madison
- Phillip Grimaldi, Rice University
- Jonathan Grudin, Microsoft Research
- Sumit Gulwani, Microsoft
- Philip Guo, University of Rochester
- Greg Hager, Johns Hopkins, CCC
- D. Fox Harrell, Massachussets Institute of Technology

- Peter Harsha, Computing Research Association
- Marti Hearst, University of California, Berkeley
- Sabrina Jacob, Computing Research Association
- Mike Jones, Indiana University, Bloomington
- Yasmin Kafai, University of Pennsylvania
- Jeffrey Karpicke, Purdue University
- Caitlin Kelleher, Washington University in St. Louis
- Anthony Kelly, National Science Foundation
- Ken Koedinger, Carnegie Mellon University
- Andrew Lan, Rice University
- Mark Liberman, University of Pennsylvania
- Mimi McClure, CSR/CNS/NSF
- Danielle McNamara, Arizona State University
- John Mitchell, Stanford University
- Mike Mozer, University of Colorado, Boulder
- Zoran Popovi, University of Washington
- Debra Richardson, University of California, Irvine/CCC
- Carolyn Rose, Carnegie Mellon University
- Beth Russell, AAAS
- Majd Sakr, Carnegie Mellon University
- Mihaela van der Schaar, University of California, Los Angeles
- Sanjit Seshia, University of California, Berkeley
- Beth Simon, Coursera
- Rishabh Singh, Microsoft
- Armando Solar-Lezama, Massachussets Institute of Technology
- Candace Thille, Stanford University
- Kevin Wilson, Knewton
- Helen Wright, Computing Community Consortium
- Jerry Zhu, University of Wisconsin, Madison
- Ben Zorn, Microsoft Research

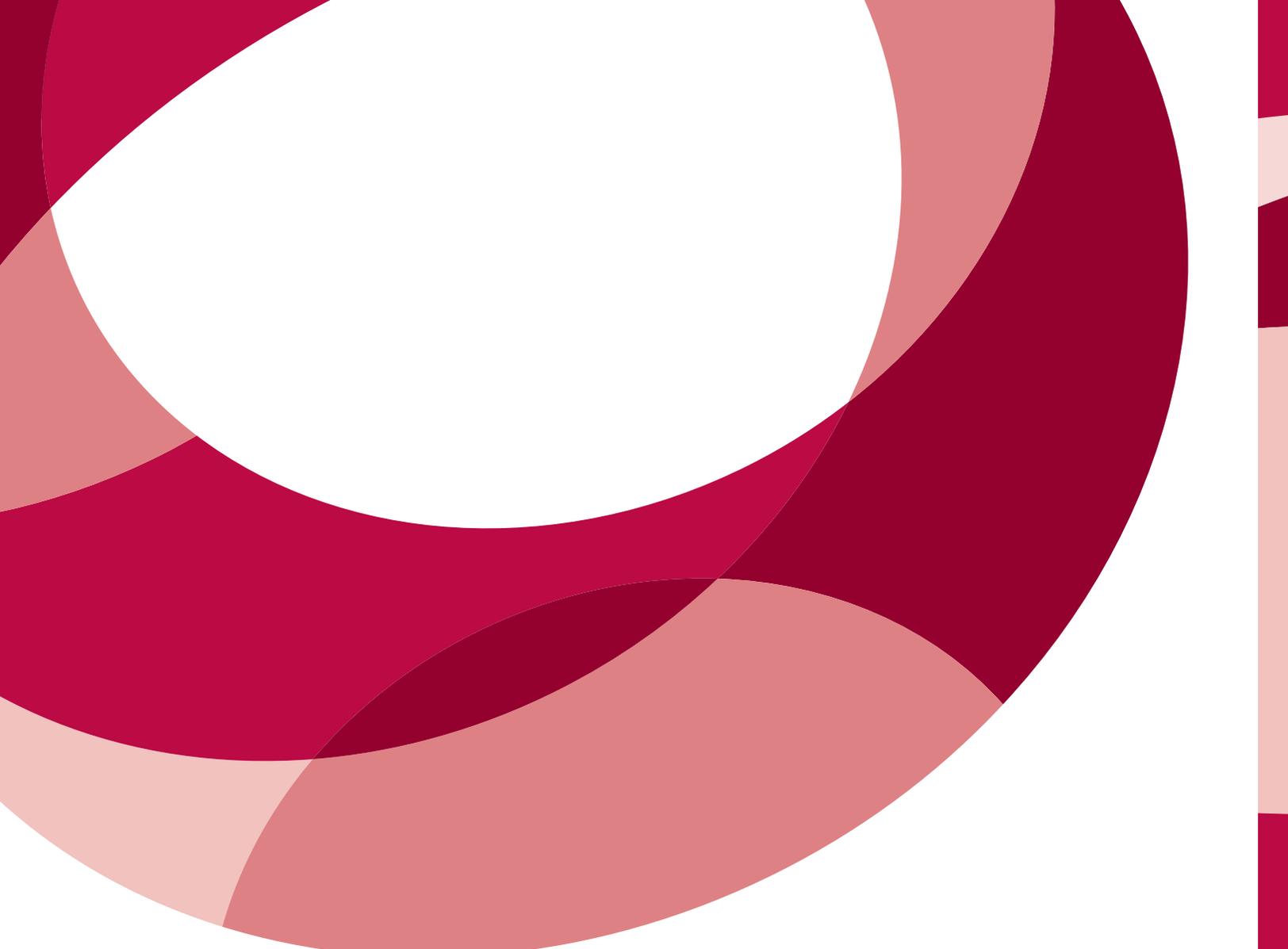

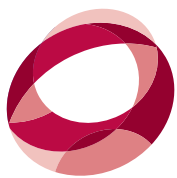
**CCC**
Computing Community Consortium
Catalyst

1828 L Street, NW, Suite 800
Washington, DC 20036
P: 202 234 2111 F: 202 667 1066
www.cra.org cccinfo@cra.org